\documentclass[10pt,journal,compsoc]{IEEEtran}                

\pdfoutput=1
\ifCLASSINFOpdf
  \usepackage[pdftex]{graphicx}
\else
\fi

\usepackage{hyperref}
\usepackage{url}

\hypersetup{
    colorlinks=true,
    citecolor=black,
    linkcolor=black,
    filecolor=black,      
    urlcolor=black,
    }

\usepackage{xcolor}
\usepackage{enumitem}
\usepackage{graphicx}
\usepackage{soul}
\usepackage{booktabs} 

\usepackage{array}
\usepackage{tikz}
\usetikzlibrary{matrix}

\tikzset{my arrow/.style={
  blue!60!black,
  -latex
  }
}

\definecolor{color1}{RGB}{229, 148, 12}
\definecolor{color2}{RGB}{56, 152, 171}
\definecolor{color3}{RGB}{166,86,40}

\newcommand{\etal}{et~al.}
\newcommand{\eg}{e.g.}
\newcommand{\ie}{i.e.}
\newcommand{\figspace}{\vspace{-7mm}}

\newcommand{\q}[1]{\textit{``#1''}}

\newcommand{\f}[1]{\texttt{#1}}

\newcommand{\name}{InfoColorizer}

\begin{document}

\title{\name{}: Interactive Recommendation of Color Palettes for Infographics}
\author{
Lin-Ping~Yuan,~Ziqi~Zhou,~Jian~Zhao,~Yiqiu~Guo,~Fan~Du,~Huamin Qu
\IEEEcompsocitemizethanks{
\IEEEcompsocthanksitem
Lin-Ping Yuan and Huamin Qu are with the Hong Kong University of Science and Technology. E-mails: \{lyuanaa, huamin\}@cse.ust.hk.

\IEEEcompsocthanksitem
Ziqi Zhou and Jian Zhao are with the University of Waterloo. E-mails: \{z229zhou, jianzhao\}@uwaterloo.ca.
\IEEEcompsocthanksitem
Yiqiu Guo is with the Xi'an Jiaotong University. Email: maxleaf@stu.xjtu.edu.cn.

\IEEEcompsocthanksitem
Fan Du is with Adobe Research. Email: fdu@adobe.com.
}}


\markboth{Under Review}
{Yuan \MakeLowercase{\textit{et al.}}: \name{}: Interactive Recommendation of Color Palettes for Infographics}

\IEEEtitleabstractindextext{
\begin{abstract}
When designing infographics, general users usually struggle with getting desired color palettes using existing infographic authoring tools, which sometimes sacrifice customizability, require design expertise, or neglect the influence of elements’ spatial arrangement. 
We propose a data-driven method that provides flexibility by considering users’ preferences, lowers the expertise barrier via automation, and tailors suggested palettes to the spatial layout of elements. 
We build a recommendation engine by utilizing deep learning techniques to characterize good color design practices from data, and further develop \name{}, a tool that allows users to obtain color palettes for their infographics in an interactive and dynamic manner. To validate our method, we conducted a comprehensive four-part evaluation, including case studies, a controlled user study, a survey study, and an interview study. The results indicate that \name{} can provide compelling palette recommendations with adequate flexibility, allowing users to effectively obtain high-quality color design for input infographics with low effort.
\end{abstract}

}

\maketitle
\IEEEdisplaynontitleabstractindextext
\IEEEpeerreviewmaketitle
\section{Introduction} 
Infographics have been widely accepted as an effective means to convey abstract information to the general public. 
Besides the content and structure of infographic elements (\eg, shapes, pictograms, text, and indices), the colors of these elements and their combination---\textit{color palette}---are essential, because it significantly influences its aesthetics, engagement, and memorability \cite{harrison2015infographic, borkin2013makes}. 
However, either crafting an infographic or selecting an effective palette is not easy, especially for general users who lack expertise in design, since each task requires considering many factors simultaneously such as layout, appearance, and perceptual effectiveness. 
While many authoring tools \cite{kim2016ddg, liu2018illustrator, xia2018dataink, wang2018infonice, chen2019towards, cui2019text, wang2019datashot} have been developed to facilitate infographics creation, these tools do not provide adequate color design supports. Users are required either to manually craft color palettes or choose them among a predefined set. 

Imagine a marketing manager, Linda, obtains a blue-background infographic online, and wants to use it in her slides with the company brand theme, which is red. 
She loads the infographic into Adobe Illustrator, but soon gets stuck in attempting to create a color palette from scratch.
While there are many principles for color design, Linda is not familiar with them, and thus has no idea how to leverage them to get a harmonious palette. 
Thus, she turns to the predefined palettes in the tool, but finds limited available choices to satisfy her needs. 
She wants the background red while having some elements' colors to reflect affective or semantic information.
Even a palette meeting all the requirements is finally found, there is still a big question on which color in the palette should be applied to which element of the infographic. 
The spatial layout of these elements matters \cite{palmer2013visual, lin2013probabilistic}, for example, a piece of text of less contrast color with its background element is hard to read.

The above example reveals three key challenges of designing color palettes using the existing tools: 1) creating a palette from scratch requires users having relevant expertise, 2) using predefined palettes by the tool limits users' freedom, and 3) applying a palette to an infographic is complicated due to the spatial layout of elements.

\begin{figure}[!tb]
    \centering
    \includegraphics[width=\linewidth]{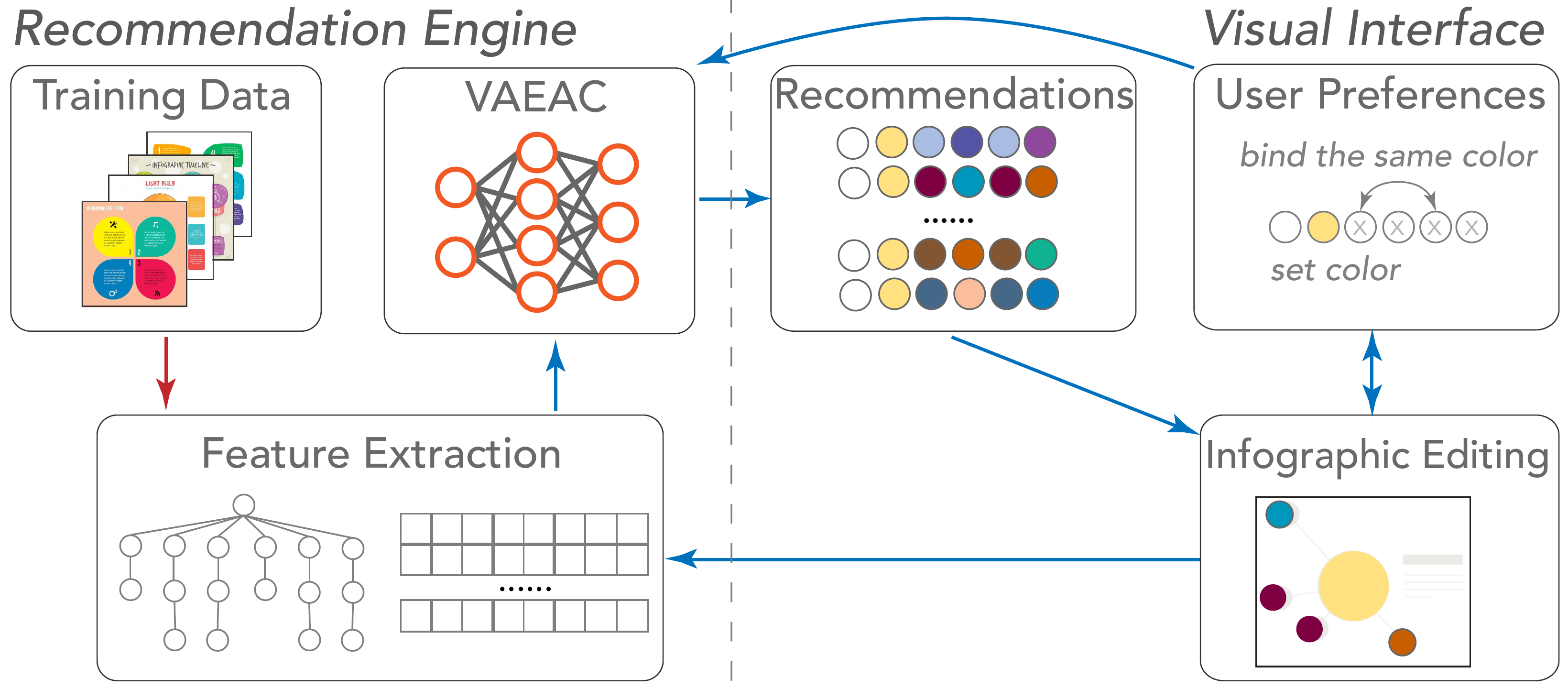}
    \vspace{-8mm}
    \caption{
    \name{} consists of a recommendation engine and a visual interface. 
    The recommendation engine is constructed by first extracting structure and color information from inforgraphics and then training a deep learning model, VAEAC, to characterize good color design practices.
    The visual interface further allows users to obtain recommended palettes, specify various preferences, edit infographics, and retrieve new recommendations iteratively. 
    }
    \label{fig:architecture}
\end{figure}

To address these challenges, we propose \textit{\name{}}, an interactive tool that allows general audience to effectively design color palettes during infographic creation, using a data-driven approach (\autoref{fig:architecture}).
We employ deep learning to extract color design practices from a large dataset of infographics created by designers, and then use the learned model to recommend appropriate color palettes. This \textit{lowers expertise barrier} of users to craft good color palettes. 
Particularly, we frame the learning process as a conditional generative problem, and leverage VAEAC (Variational AutoEncoder with Arbitrary Conditioning) \cite{ivanov2018variational} to recommend color palettes dynamically based on conditions (\eg, color preferences) set by users. This \textit{offers flexibility} to users by enabling partial specification of palettes with exact or vague color constraints. 
Moreover, we characterize infographics with features including the information of element spatial layouts in the dataset, allowing for integrating such knowledge into our learned model. This suggests color palettes \textit{tailored for particular element arrangements} in infographics.
\name{} also supports some basic editing functions, allowing users to try out different infographic layouts, obtain corresponding color palette recommendation, and iteratively refine their design.
The source code of the system (including the models, user interface, and examples) will be available at ~\url{https://github.com/yuanlinping/InfoColorizer}.

We validated \name{} through a comprehensive evaluation containing four parts.
First, we demonstrate the usefulness of \name{} with case studies using real-world infographics and example scenarios. 
These cases reveal that the system can facilitate color palettes design with cogent recommendations in different tasks such as filling empty wireframes and combining infographics with different color schemes.
Then, we conducted a controlled user study with 24 design novices. The qualitative and quantitative results show that \name{} offers higher efficiency and better creativity support than a baseline with manual color design with online resources.
Third, we carried out an online survey study with 102 users to compare artist-designed, predefined, randomly-generated, baseline-crafted and \name{}-recommended color palettes on aesthetics and readability.
The results indicate that although \name{}'s recommendations were not perceived as good as artist-designed palettes, they received higher scores than the other three methods on both factors.
Finally, we interviewed four graphic design experts in depth;
they appreciated \name{}'s novel features and were able to generate compelling infographics meeting their needs effectively within a few operations.
In summary, our main contributions include:
\begin{itemize}[leftmargin=1.0em,nosep]
    \item A novel data-driven approach that recommends palettes for infographics by leveraging deep learning techniques with the consideration of elements' spatial arrangements, while offering flexibility for user preferences of colors;
    \item An interactive tool, \name{}, that incorporates the data-driven recommendation and makes it easily accessible and manageable to users, along with the support of iterative design and basic infographic editing; and
    \item Insights and results from a series of evaluations covering case studies, a controlled user study, an online survey, and an interview study.
\end{itemize}

\section{Related Work}

\subsection{Infographic Models and Authoring Tools}

Compared to plain charts or text, infographics incorporate additional visual embellishments, such as icons, shapes, images, and pictograms, to convey abstract information appealingly.
Previous empirical studies have shown that embellished charts and pictographics increase comprehension, engagement, and memorability, while not reducing viewers' accuracy~\cite{bateman2010useful,haroz2015isotype}.
Further, researchers have demonstrated that colors are essential to make visualizations memorable and influence a first impression~\cite{borkin2013makes,harrison2015infographic}.

Besides empirical studies, data-driven methods or machine learning models were applied for tagging visual and textual elements in infographics~\cite{bylinskii2017understanding}, estimating the visual importance of elements~\cite{bylinskii2017learning} based on crowdsourcing data~\cite{kim2017bubbleview}, exploring perceived personalities~\cite{zhao2018characterizes}, and computing similarity of infographics to facilitate search~\cite{saleh2015learning}.
Recently, Lu \etal~\cite{lu2020exploring} explored high-level narrative flows of infographics extracted from online collections. 
However, none of the above works focuses on the color aspect of infographics design.
Inspired by these techniques, we employ a data-driven method to extract good color design practices and utilize them for recommendations.

To facilitate the creation of an aesthetic infographic, researchers have developed many authoring tools, which fall into three main categories: programming toolkits, interactive design environments, and semi- or fully-automated tools.
Programming toolkits (\eg, D3~\cite{bostock2011d3}) allow users to create visualizations at the greatest extent of flexibility. 
However, they usually have a steep learning curve and are challenging for general users. 
Thus, many interactive design environments have been developed to facilitate users without programming skills, such as Adobe Illustrator. 
Also, tools that support data binding were proposed to ease the creation of data-driven charts and infographics, including the binding of standard marks~\cite{satyanarayan2014lyra}, icons~\cite{wang2018infonice}, and hand-drawn shapes~\cite{kim2016ddg, xia2018dataink,liu2018illustrator}.
However, they still require users to manually craft infographic elements.
To further lower the barrier, semi- or fully-automatic tools were developed to support creating infographics directly from different inputs, such as natural language statements~\cite{cui2019text}, tables~\cite{wang2019datashot}, and timeline images~\cite{chen2019towards}. 

While providing rich capabilities, almost all the above systems leave users with two options to obtain a color palette---which is critical for the aesthetics of their infographics---including: 1) manually creating one from scratch and 2) choosing from a predefined collection. 
The former requires expertise in color design and the latter limits a user's flexibility. 
Our work aims to address these issues via interactive recommendation.
Motivated by the discussion of Lu~\etal~\cite{lu2020exploring} on the spatial structure of infographics, we go a step further to integrate the information of elements layout into suggesting proper color palettes. 

\subsection{Color Palette Design Principles and Tools}
Prior studies mainly focus on improving perceptual discriminability and aesthetics of designed color palettes in data visualization. 
Levkowitz~\etal~\cite{levkowitz1992color} suggested that discriminable palettes should effectively mirror encoded data and accurately convey data differences. 
Visualization designers are recommended to consider many factors, such as underlying data types~\cite{tennekes2014tree}, visual consistency~\cite{qu2018keeping}, tasks~\cite{tominski2008task}, and color properties (from basic visual separability~\cite{wang2018optimizing}, contrast~\cite{mittelstadt2014methods}, and difference metrics~\cite{szafir2018modeling}, to more advanced color appearance~\cite{lee2012perceptually}, name differences~\cite{heer2012color}, affect~\cite{bartram2017affective}, and semantics~\cite{lin2013selecting, setlur2015linguistic}).
Moreover, studies indicated that using harmonious colors~\cite{meier2004interactive} or adjusting hue and saturation~\cite{moreland2009diverging} can increase the aesthetics of visualizations.

However, applying these principles in practice can be difficult for non-experts.
Many techniques have been proposed to ease this process, such as ColorBrewer~\cite{harrower2003colorbrewer} that provides predefined color palettes for encoding sequential, diverging, and qualitative data. 
In the case of graphic design (\ie, not encoding any data), some commercial interactive tools (\eg, Adobe Color~\cite{adobecolor2020}) utilize harmonic templates to help users craft customized color palettes with high quality. 

Further, researchers have developed algorithms to automatically generate color palettes for different applications.
For categorical data, Colorgorical~\cite{gramazio2016colorgorical} creates palettes based on user-defined balance of color discriminability vs. aesthetic preference.
Palettailor~\cite{lu2020palettailor} further provides a data-aware approach that generates and assigns colors for multi-class scatterplots, bar and line charts to maximize their visual discrimination. 
For numerical data, Smart~\etal~\cite{smart2019color} modeled expert-designed color ramps to suggest palettes.
Besides data charts or visualizations, algorithms have been designed for (re)coloring other types of media, such as natural images and patterns~\cite{cohen2006color,nguyen2017group}. 
O'Donovan \etal~\cite{odonovan2011color} proposed a color compatibility model that can score the quality of any five-color palettes.
By considering compatibility and spatial arrangements of colors, Lin \etal ~\cite{lin2013probabilistic} suggested a probabilistic model for coloring 2D patterns.

However, the above techniques focus on data charts/ visualizations, natural images, or patterns; none of them are developed for infographics with unique characteristics. 
First, infographics consist of both data elements and visual embellishments, where colors play multiple roles such as visual group indicators and artistic decorators~\cite{lu2020exploring}. 
Second, the spatial arrangement of its elements is different from that in other media, which may exhibit complicated semantics and convey a narrative. 
Perhaps the general methods on suggesting palettes for website design~\cite{adobecolor2020, odonovan2011color, harrower2003colorbrewer,gramazio2016colorgorical} can be used for infographics. 
But they are limited in generating palettes with a fixed number of colors (\eg, five), and do not indicate how users assign colors to infographic elements.

\subsection{Visualization Recommendation}
Researchers have explored various techniques for recommending appropriate visualizations, including some commercial tools such as Tableau.
One category falls in rule-based methods.
APT~\cite{mackinlay1986automating} introduces a compositional algebra to enumerate the space of charts and ranks them, which was later extended in SAGE~\cite{roth1994interactive}.
CompassQL~\cite{wongsuphasawat2016towards}, the basis of Voyager~\cite{wongsuphasawat2016voyager} and Voyager 2~\cite{wongsuphasawat2017voyager}, offers flexible query specifications for searching the visualization space and providing recommendations. 
Further, Draco~\cite{moritz2019formalizing} leverages answer set programming to describe constraints over visualization design.
Another category is data-driven, based on machine learning techniques. 
VizML~\cite{hu2019vizml} learns design choices from a corpus of data vs. visualization pairs. 
Data2Vis~\cite{dibia2018data2vis} is an end-to-end neural network that generates visualizations directly from data.
DeepEye~\cite{luo2018deepeye}, on the other hand, combines rule-based methods and machine learning to rank and classify visualizations.

While the above systems allow users to effectively create visualizations from input data, none of them adequately supports recommending color designs of generated charts.
Moreover, as mentioned above, infographics have unique characteristics that are different from ordinary charts or visualizations, which is the focus of our work. 
\section{\name{} Design and Overview}
In this section, we outline the design goals for developing \name{}, followed by an overview of our method.

\subsection{Design Goals}
As most infographics tools focus on authoring the geometrical content of infographic elements, our main goal is to facilitate the essential follow-up step---color design. 
Motivated by the aforementioned scenario and limitations of the existing tools, we derive the following design goals to guide the development of \name{}. 

\textbf{G1: Lower expertise barrier for crafting professional color palettes.} 
Graphic designers consider many factors simultaneously when creating high-quality palettes, such as aesthetics, harmony, and perceptual discriminability. 
However, this task is challenging for general users due to the lack of expertise, as there could be unlimited numbers of choices for a color palette. 
\textit{The system should characterize good practices embodied in handcrafted designs, and assist users with automatic palette recommendation that reflects these good practices.} 

\textbf{G2: Offer flexibility to embed different kinds of user preferences.} 
Users may have preferences and constraints when creating palettes. 
For example, they may want to apply a specific or semantically meaningful color to a particular element. 
They may also want to set multiple relevant elements with the same color for consistency.
\textit{The system should provide a flexible mechanism to allow for specifying various types of color preferences on elements of interest, as well as alleviate users from considering colors for other elements.}
    
\textbf{G3: Incorporate consideration of spatial arrangements of elements.} 
The spatial layout of elements in an infographic greatly influences its perceived color appearance.  
Even if a palette looks good independently, it can have poor performance after being applied to an infographic. 
Further, there exist numerous ways to apply a palette to an infographic (\eg, around $5^{10}$ assignments for a five-color palette and a ten-element infographic), causing much trial-and-error tweaking.
\textit{The system should adapt palettes to particular spatial arrangements of input infographics in recommendations, thus freeing users from the tediousness of tuning color assignments.}

\textbf{G4: Support simple user interactions and iterative design of color palettes.} 
General users rely on an easy interface for accessing different system functions.
\textit{The system should provide intuitive user interactions such as obtaining effective palette recommendation, specifying color preferences and other constraints, and previewing \& editing infographics.} 
Further, the color palette design process is often iterative by trying different ideas. 
\textit{The system should facilitate refining results in a human-in-the-loop manner, such as bookmarking recommended palettes, storing history of recommendation, and tuning constraints according to their needs.}

\subsection{Method Overview} \label{sec:method-overview}

Based on the above goals, we develop \name{}, a visual system that provides interactive palette recommendation for an infographic with flexible user preference settings. 
As shown in \autoref{fig:architecture}, we employ a data-driven approach to automatically acquire good practices exhibited in infographic collections and then utilize the ``learned knowledge'' to recommend palettes, with a visual interface that allows user interactions with the underlying recommendation.

More specifically, considering an infographic $\mathbf{I}=\{E_1, E_2, \cdots, E_n\}$, where $E_i$ is an element, we characterize $\mathbf{I}$ with a set of non-color features $\mathbf{F}=\{F_1,F_2,\cdots,F_m\}$ and color features $\mathbf{C}=\{C_1, C_2, \cdots, C_n\}$ for the $n$ elements (see \autoref{sec:infographic-model}). 
The non-color features $\mathbf{F}$ include information at different granularity (\eg, infographic and element levels), and the spatial arrangement of elements, which are combined and represented in a tree structure (G3).
For expert-designed infographics, the color features $\mathbf{C}$, and their relations with the features $\mathbf{F}$ reflect good practices that we wish to capture. 

We therefore frame our recommendation process as a conditional generative problem (see \autoref{sec:system}).
We employ Variational AutoEncoder with Arbitrary Conditioning (VAEAC) ~\cite{ivanov2018variational} as our generative model, because of its flexibility in adapting any features as conditions. 
That is, given a collection of expert-designed infographics, with features $(\mathbf{F}_k,\mathbf{C}_k)$, the model can learn a probability distribution over the full feature set---non-color features $\mathbf{F}$ and colors $\mathbf{C}$---to capture the good practices (G1). 
Later, the learned model can be used to generate any ``missing'' features of an infographic $\mathbf{I}$ with knowing the rest (\ie, the arbitrary conditions).
For example, users can specify colors $C_i$ and $C_j$ for certain elements $E_i$ and $E_j$, and the conditional generative problem becomes sampling from $p(\mathbf{C}\setminus C_{i,j} | \mathbf{F},C_{i,j})$, allowing for the flexibility of incorporating different kinds of user preferences (G2). 
To make the above recommendation easily accessible and configurable, we design a visual interface for \name{}, which also enables iterative generation of colors and simple infographic editing functions (G4).

\section{Dataset and Infographic Model} \label{sec:infographic-model}
To achieve the design goals, the starting step is to identify high-quality infographic datasets from which a data-driven method can extract good palette design practices (G1).
Further, we need to conceptually model infographics in a form that is effective for algorithms to understand and process. 
In this section, we discuss the above two aspects.

\subsection{Dataset}

Previous studies collected several infographic datasets, such as MassVis~\cite{borkin2013makes,massvis}, Visually29K~\cite{bylinskii2017understanding, bylinskii2017learning,visually29k}, InfoVIF~\cite{lu2020exploring,infovif} and Timelines~\cite{chen2019towards,timelines}.
In this work, we chose InfoVIF (containing 13,245 infographics) as our initial test bed for the following reasons. 
First, compared to MassVis and Visually29K, InfoVIF tends to be more useful for general audiences, because most items are design templates that can be used as a starting point to create personalized infographics.
Second, InfoVIF contains infographics with more uniform styles of visual elements and layouts than those in MassVis and Visually29K, allowing machine learning to better capture common design patterns in infographics.
Third, compared to Timelines, InfoVIF has a broader coverage of infographics, including not only timelines but also other types.
Finally, infographics in InfoVIF are contributed by world-wide designers with high-quality and diverse design themes. 
Thus, InfoVIF is a suitable resource from which good color design practices can be extracted (G1).

\subsection{Conceptual Model of Infographics} \label{sec:conceptual-model}

\begin{figure}[!tb]
    \centering
    \includegraphics[width=\linewidth]{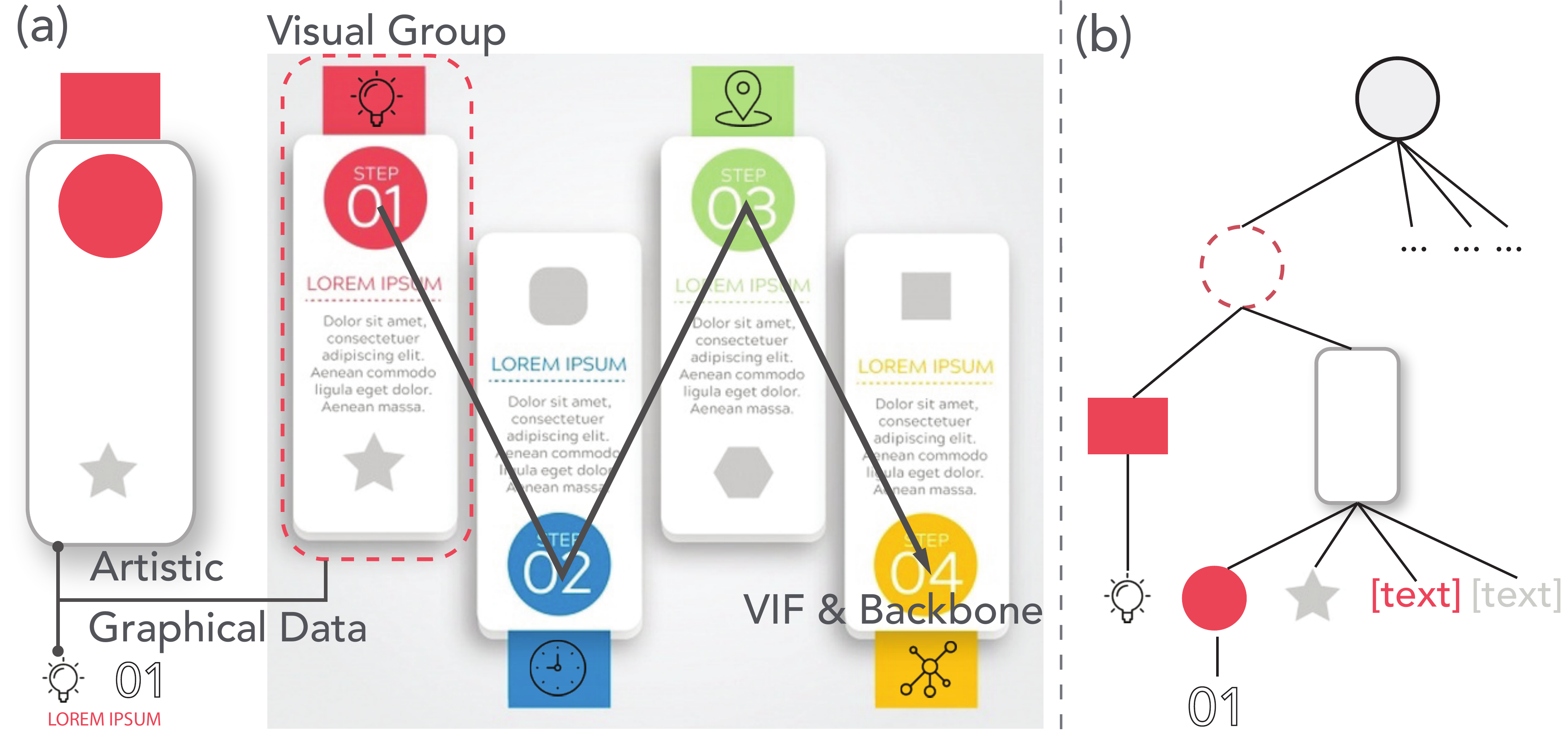}
    \vspace{-8mm}
    \caption{(a) Artistic elements, graphical data elements, visual groups, and visual information flow (VIF) \& backbone in an infographic. (b) The conceptual tree model. Some branches are omitted for simplicity.
    }
    \label{fig:infographic_model}
\end{figure}

As shown in \autoref{fig:infographic_model}-a, Lu~\etal~\cite{lu2020exploring} characterized an infographic as a combination of \textit{artistic decorations} (\eg, shapes, images, and cliparts) and \textit{graphical data elements} (\eg, icons, text, and indices).
Graphical data elements are then organized into \textit{visual groups} to convey pieces of information.
Visual groups are further presented in sequential order, called \textit{visual information flow (VIF)}, to indicate a narrative of the infographic. 
The path connecting the semantic centroids of the visual groups is called VIF \textit{backbone}.

However, this infographic model is not sufficient for our case, because it \textit{only} considers graphical data elements when composing visual groups.
This is oversimplified, because the artistic decorations and their colorfulness largely affect the aesthetics of infographics, and thus are important in determining proper color palettes for a visually compelling infographic.
Further, while VIF captures a high-level spatial structure to make logical sense, more detailed spatial relationships between individual elements arguably influence the color palette design.
For example, two elements next to each other (\ie, \textit{adjacency}) may share the same color to imply the ``Similarity'' Gestalt Law \cite{desolneux2004gestalt}; and one element fully on top of another (\ie, \textit{inclusion}) is benefited from choosing a high contrast color with respect to the one below. 

To address these issues, we first extend Lu \etal's model \cite{lu2020exploring} by including both artistic and graphical data elements in visual groups, as well as VIFs, which characterizes an infographic from a logical perspective (\autoref{fig:infographic_model}-a). 
Inspired by trees being used to analyze topological structures of images~\cite{suzuki1985topological}, we further employ a tree structure, which incorporates Lu \etal's model and characterizes an infographic from a spatial perspective (G3).
The structure can describe the adjacency and inclusion spatial relationships at multiple levels (\autoref{fig:infographic_model}-b).
The root node of a tree represents the whole infographic (\ie, the background canvas), and the second layer of the tree represents all the visual groups, whose descendant nodes are artistic or graphical data elements within the corresponding visual group. 
Under a visual group node, a parent-child link indicates element inclusion, and a sibling relation indicates adjacency in layout.
Our model conceptualizes an infographic from both logical and spatial aspects, allowing data-driven models to extract associations between these aspects and color designs.

\section{\name{} System} \label{sec:system}

In this section, we describe \name{} in detail. As shown in \autoref{fig:architecture}, the system consists of a recommendation engine and a visual interface. 
We first describe the feature extraction process (\autoref{sec:features}) and how we characterize good color design practices using the VAEAC model (\autoref{sec:recommendation-model}).
We then demonstrate \name{}'s ability to support flexible color preferences specification (\autoref{sec:user-preferences}), and the visual interface that enables an effective palette design workflow (\autoref{sec:interface}).

\subsection{Characterize Infographics with Various Features} \label{sec:features}

For an infographic $\mathbf{I}$, we compile a set of color $\mathbf{C}$ and non-color $\mathbf{F}$ features to describe its visual appearance and spatial arrangement of elements (G3).

We extract the color values of all elements and the infographic background as the color features $\mathbf{C}$. CIELab color space is used because of its perceptual uniformity. 
We distill a list of non-color features $\mathbf{F}$ at the multiple levels. 
At \textbf{infographic level}, we obtain \textit{VIF Type}, \textit{Visual Group Number}, and \textit{Visual Group Distance}.
Features in \textbf{visual group level} contain \textit{Visual Group Element Number} and \textit{Relative Visual Group Size}.
At \textbf{element level}, we extract \textit{Element Type}, \textit{Relative Element Size} and \textit{Relative Element Pixel Area} for each artistic and graphical data element. 
In addition, we use the nested set model \cite{feichtner2005topology} to represent the \textbf{tree structure} (\autoref{sec:conceptual-model}) of an infographic by storing \textit{Left Index Number and Right Index Number} of each node. 
A complete explanation of the above features and technical details about the extraction algorithms can be found in Appendix Sec. A.

The non-color features embed many design factors considered by experts. Specifically, they help organize information about narrative flow (\textit{VIF}), visual groups (\textit{Visual Group Number and Distance}), size(\textit{Relative Visual Group Size, Relative Element Size/Pixel Area}), shape (\textit{Element Type}).
Besides, the tree structure reveals the spatial arrangement of elements (G3) and other design factors, such as layers by tree traversal background/foreground contrast by inferring parent-child relationships, etc.
Characterizing these factors makes capturing high-level color design practice from infographics possible with a data-driven approach.

\subsection{Lower Expertise Barrier with Recommendation} \label{sec:recommendation-model}
The next step is to train a machine learning model that extracts good design practices in data and utilizes that for providing recommendations (G1).
As mentioned in \autoref{sec:method-overview}, we frame the recommendation process as a conditional generative problem, and employ Variational AutoEncoder with Arbitrary Conditioning (VAEAC) \cite{ivanov2018variational}, a special kind of \textit{VAE} (Variational AutoEncoder) \cite{kingma2014auto}.

In general, a VAE aims to learn a \textit{bidirectional} mapping between data items $x$ in an application domain and variables $z$ in a continuous latent space.
The model consists of two parts---an encoder $p(z|x)$ that converts $x$ to $z$ and a decoder $q(x|z)$ that does the opposite; and the training process is to learn the two probability distributions. 
In our case, $x$ is the infographic features, $[\mathbf{F},\mathbf{C}]$, and $z$ reflects the abstract knowledge in data.

We want to obtain color palettes according to a specific design of infographic structure reflected in $\mathbf{F}$, which is a \textit{conditional} generation. 
Moreover, users may have specific preferences on coloring certain infographic elements (G2), meaning that some part of $\mathbf{C}$ is in the conditions.
VAEAC, as an extension of VAE, can handle the above requirements, by incorporating a binary mask vector $b$ that controls which part of $x$ is the conditions (\textit{observed} features) or not (\textit{unobserved} features).
Its encoder and decoder are $p(z|x_{1-b},b)$ and $q(x_b|z,x_{1-b},b)$, respectively, where $x_{1-b}$ is the observed part and $x_b$ is the unobserved part.
The model is trained with a full set of features $x$, and can fill in any ``missing'' unobserved part of $x$ (\ie, called \textit{feature imputation}) during the generation stage with a dynamic input of $b$.
When applying to our case, we treat $\mathbf{F}$ always observed and set part of $\mathbf{C}$ observed, controlled by a user input $b$.
If none of $\mathbf{C}$ is observed (\ie, $b=\vec{1}$), the model can generate a full color palette for an infographic; and if some colors of $\mathbf{C}$ is observed (\ie, specified by a user with certain colors), the model can generate the rest of a color palette with these colors satisfying users' constraints. 
Details regarding the user preferences support will be introduced in \autoref{sec:user-preferences}.

We trained VAEAC using the obtained features for infographics, $[\mathbf{F},\mathbf{C}]$, described in \autoref{sec:features}.
We also considered two alternative models including: 1) GAIN (Generative Adversarial Imputation Nets) \cite{pmlrv80yoon18a}, which is the state-of-the-art GAN-based model for feature imputation, and 2) MICE (Multivariate Imputation by Chained Equations) \cite{buuren2011mice}, which is a classic non-deep-learning method. 
Our experiments indicated that VAEAC performed the best on the dataset. Details about the model training, comparison, and evaluation can be found in Appendix Sec. B.

\subsection{Offer Flexibility with Versatile User Preferences} \label{sec:user-preferences}
With a trained VAEAC model, \name{} can not only recommend colors for each infographic element but also support flexible control over the colors in recommendation (G2).
\name{} supports two kinds of user constraints: 1) specifying colors for certain elements in either an exact or a vague form, and 2) binding several elements together to have the same color in recommended palettes.

As mentioned in \autoref{sec:recommendation-model}, VAEAC can generate unobserved colors conditioning on non-color features and observed colors. 
Thus, we can generate palettes meeting users' preferences by manipulating the input feature vector. 
Specifically, if a user assigns an exact color (\eg, in CIELab space $[l,a,b]$) to an element $E_i$, the corresponding color features of $E_i$ are set to $[l,a,b]$, indicating these features are observed.
Moreover, users can assign colors to an element $E_i$ semantically using a word (referring to a range of colors) such as a \textit{color name} (\eg, red, skyblue), an \textit{object} (\eg, apple, dollar), and a kind of \textit{affect} (\eg, exciting, calm).
To handle such vague specifications, we first collected over 200 \textit{(word, colors)} pairs from the previous works on color names \cite{heer2012color}, color affect \cite{bartram2017affective}, and color semantic \cite{lin2013selecting}, as well as a website \cite{emotionwheel}. 
We then utilize this information to manipulate the input feature vectors. 
When a word is assigned to an element $E_i$, we randomly select $k$ colors from the corresponding color set and then generate $k$ input vectors (we set $k$ to 3 in \name{}), where each has a different color (from $[l,a,b]_1$ to $[l,a,b]_k$) for the observed features of $E_i$. 
We can thus obtain $k$ sets of recommendations and randomly pick some for presentation.

Users can also bind relevant elements (\eg, those within a visual group, or all icons, text, etc.) to constrain them with the same color in recommendations.
We adopt a post-processing method on the recommended color palettes.
For example, suppose that $E_i,E_j,E_k$ are bound, for each recommended color palette, we randomly select one of them based on a probability decided by their areas, and then set all three elements with the color of the selected one.

\subsection{Support User Workflow with Visual Interface} \label{sec:interface}

We develop a visual interface that enables users to iteratively obtain desired palettes by supporting basic editing and previewing functions, color preference and other constraint specification, and interactive recommendation (G4).

\begin{figure*}[tb]
  \centering
  \includegraphics[width=\linewidth]{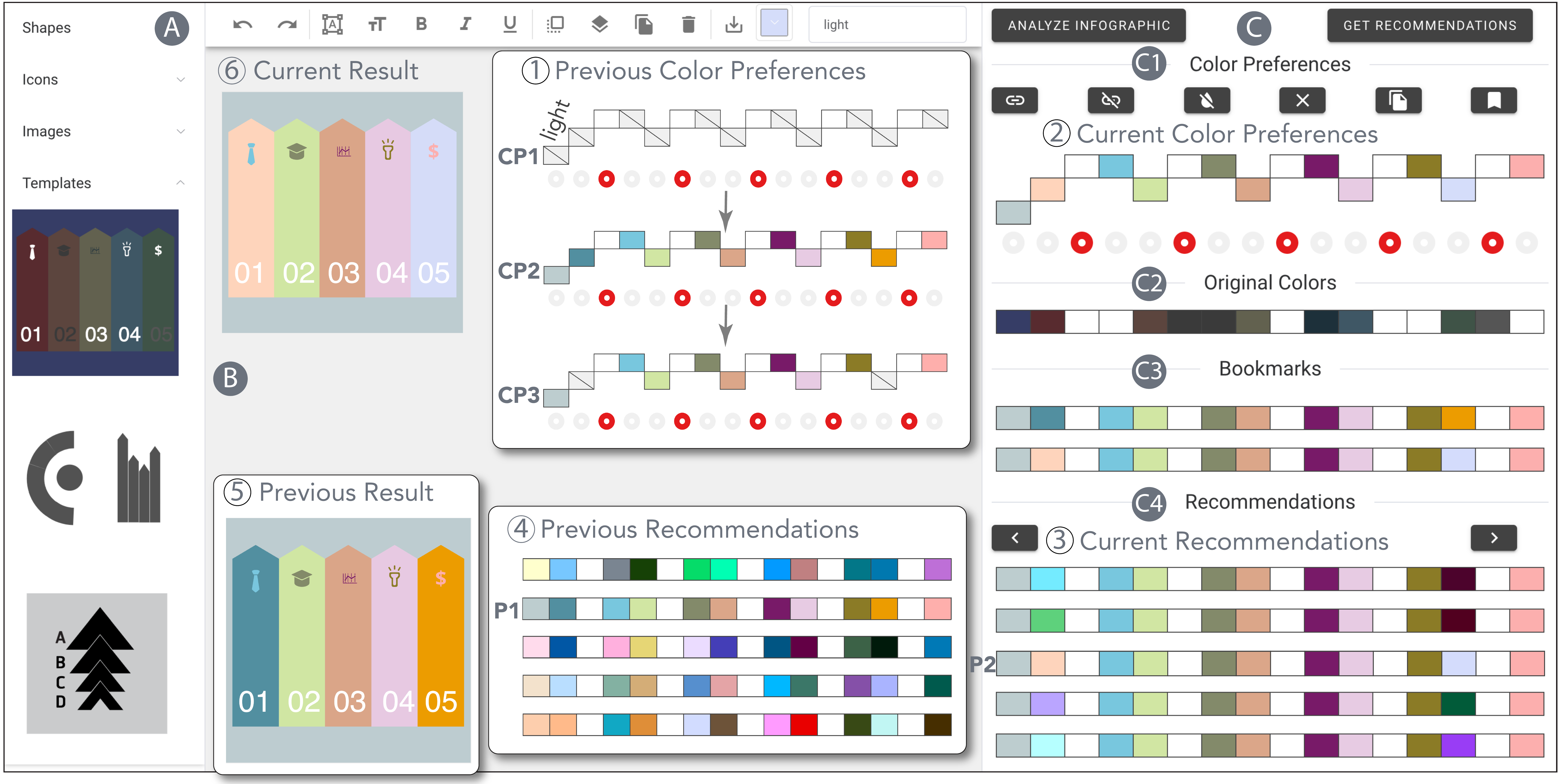}
  \figspace{}
  \caption{\name{} consists of three linked panels: (A) Content Library stores materials for creating infographics; (B) Main Canvas supports simple editing functions to compose infographics; (C) Content Panel offers setting color preferences and viewing recommendations, composed of four sections: (C1) Color Preferences, (C2) Original Colors, (C3) Bookmarks, and (C4) Recommendations. It allows users to obtain desired color palettes by iteratively \textcircled{1}\textcircled{2} specifying preferences, \textcircled{3}\textcircled{4} getting recommendations, and \textcircled{5}\textcircled{6} previewing the results for further refinement.
  }
  \label{fig:interface}
\end{figure*}

The interface (\autoref{fig:interface}) consists of three interactively-coordinated panels.
The \textit{Content Library} (\autoref{fig:interface}-A) stores raw materials (\eg, shapes, images, icons) and infographic templates.
Users can create an infographic either from scratch or based on a template, and color it using \name{}'s palette recommendation.
Users can also upload bitmap image infographics or add text, and modify the colors according to their needs. 
Selected resources can be edited on the \textit{Main Canvas} (\autoref{fig:interface}-B). The toolbar on the top supports some simple editing functions such as arrange, group/ungroup, duplicate, and delete.
The \textit{Control Panel} (\autoref{fig:interface}-C) is a core component, where users can obtain desired color palettes by iteratively specifying preferences, obtaining recommendations, and refining the design. 
Overall, we designed the interface with common panels and components to improve the usability and learnability.
However, we proposed a novel widget for setting color preferences (\autoref{fig:interface}-C1) with visualizations of layered elements and interactive linking.

Here, we use a simple scenario to demonstrate an interactive workflow of obtaining desired palettes with \name{}. 
Suppose the marketing manager, Linda, wants to improve a chosen infographic from the Content Library (the first one in \autoref{fig:interface}-A). 
She then clicks ``Analyze Infographic'' which analyzes the spatial relationships of its elements and displays a tree structure on the Color Preferences section (\autoref{fig:interface}-C1).
The tree, informing our conceptual infographic model (see \autoref{fig:infographic_model}), is visualized as rectangles in horizontal layers.
The bottom layer is the background canvas, and the second bottom layer contains the elements directly placed on the background, and so forth.
The rectangle color indicates users' preference of the corresponding element; an empty rectangle with a diagonal line means no constraint. 

Initially, she wants the background to be lighter and all the text to be pure white consistently. 
With the Color Preferences section (\autoref{fig:interface}-C1), she assigns a vague color specification with the word ``light'' to background canvas; the word is shown on top of the rectangle. 
She then binds the four text objects and sets ``\#FFFFFF'' (\ie, white) as their colors. 
The bound elements, which will always have the same colors in recommendations, are indicated by the small red dots below.
The resulting color preference setting is shown in \autoref{fig:interface}-\textcircled{1}-CP1.

Linda clicks ``Get Recommendations'', and a list of recommended palettes meeting her needs are then returned by \name{} (\autoref{fig:interface}-\textcircled{4}).
The number of returned recommendations can be adjusted in \name{} (the default is five).
She picks her favorite one, \autoref{fig:interface}-\textcircled{4}-P1, for preview and refinement.
The chosen palette is then duplicated in the Color Preferences section (\autoref{fig:interface}-\textcircled{1}-CP2), and the infographic is automatically colored by the palette (\autoref{fig:interface}-\textcircled{5}). 

However, Linda is not satisfied with the colors of the first and last bars (\ie, ``01'' and ``05'').
She thus clears the colors of the two bars (\autoref{fig:interface}-\textcircled{1}-CP3), and requests new recommendations with this preference setting. 
The results are shown in \autoref{fig:interface}-\textcircled{3}.
Similar to the previous iteration, she picks her favourite palette in \autoref{fig:interface}-\textcircled{3}-P2, which updates the Color Preference section (\autoref{fig:interface}-\textcircled{2}) and the infographic (\autoref{fig:interface}-\textcircled{6}).
Linda is quite happy with this color design and exports the infographic for her presentation slides.
\section{Evaluation}
To assess the effectiveness and usefulness of \name{}, we conducted a four-part evaluation. 
We first use several case studies to demonstrate that \name{} is able to generate compelling color palettes under different scenarios. 
In addition, we quantitatively and qualitatively evaluate \name{} from the perspectives of novice creators, infographic readers, and graphical design experts separately by conducting a controlled user study, a survey study, and an interview study.
These studies comprehensively reflect the strengths and weaknesses of \name{} on different aspects.
Detailed information about our studies can also be found in the supplementary materials.

\subsection{Case Studies}
\autoref{fig:case_study} presents a set of infographics colored by recommended palettes. More diverse and complicated cases can be found in our supplemental materials.
We demonstrate how \name{} can facilitate palette creation under different user preferences and constraints. 
We consider three use cases: a) colorizing a \textit{wireframe} infographic, b) improving the color \textit{readability} of an infographic, c) \textit{stitching} two infographics with different color schemes.
For each case, we select one infographic from a website \cite{svginfographic}. 
We demonstrate \name{}'s recommendations under four conditions: 1) \textit{no preferences}, 2) \textit{exact color} specification, 3) \textit{vague color} specifications, 4) \textit{elements binding}. 
In \autoref{fig:case_study}, we use ``Pin'' icons to indicate elements that are specified with exact colors, annotate the words on elements that are specified vaguely, and add links to elements that are bound together.
We can see that \name{} can generate compelling palettes for the source infographics under different conditions; all of the results are obtained through one to two requests of recommendations. 

\begin{figure}[!tb]
    \centering
    \includegraphics[width=\linewidth]{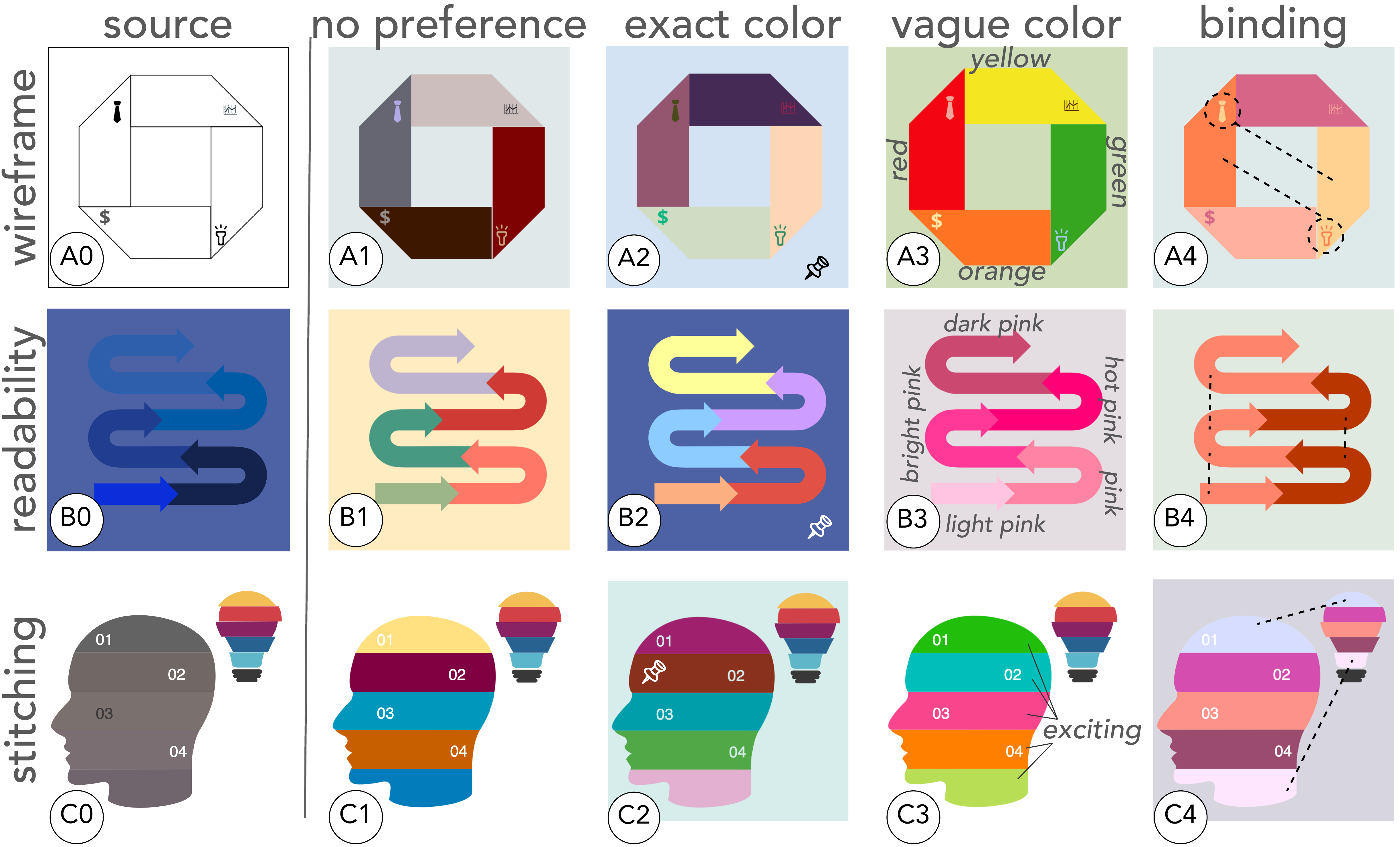}
    \vspace{-8mm}
    \caption{Example cases of three different infographic creation scenarios with four different user preference conditions.
    }
    \label{fig:case_study}
\end{figure}

Even without any color preferences, \name{} is able to suggest cogent color palettes. 
For example, in \autoref{fig:case_study}-A1, the colors of the four shapes are in a smooth and consistent theme.
In \autoref{fig:case_study}-C1, a user wants to obtain a palette for the head adapted to the bulb, and thus sets no preferences for the head while fixing colors for the bulb.
The system recommends a color palette for the head similar to that of the bulb, indicating that the model utilizes \textit{observed} color features (\ie, the bulb) to influence \textit{unobserved)} color features (\ie, the head).
Moreover, by assigning the backgrounds in \autoref{fig:case_study}-A2,B2 with desired colors, the original foreground colors are filled or enhanced with colors in contrast to the backgrounds.
Further, after specified with some words of general color hues and semantics, \name{} returns athletically pleasing infographics.
For example, in \autoref{fig:case_study}-B3, a user demands a pink-themed infographic but has no idea about what specific pink colors are the best, and thus specifies general color categories with words.
In \autoref{fig:case_study}-C3, ``exciting'' is attached to all the color strips on the head to match the underlying semantics of getting an idea (\ie, the bulb).
Finally, by binding some elements together, a user can obtain more consistent color palettes, such as linking the corresponding arrows in \autoref{fig:case_study}-B4 to indicate different types of stages, and associating the relevant parts between the head and the bulb in \autoref{fig:case_study}-C4.

\subsection{Controlled User Study} \label{sec:controlled_study}
We further conducted a controlled study to evaluate \name{} with real users on color palette design tasks.
This study aimed to investigate two aspects of the workflow:
\textbf{(S1)} whether \name{} can facilitate users for obtaining satisfactory palettes for their infographics,
and \textbf{(S2)} whether \name{} can support users' creativity in designing palettes.
In particular, we investigate these questions under the situations that users have specific color preferences and consider the spatial layout of graphic elements. 

\subsubsection{Study Setup}
\textbf{Baseline.}
To better assess the pros and cons of \name{}'s novel features, we considered a baseline to simulate a common color palette design process in practice, in which people derive palettes from different sources, including color pickers, predefined palettes, and online websites~\cite{jalal2015portraits}, and then color their infographics with a design tool.
We thus developed a baseline version of \name{} by disabling the palette recommendation function and allowing users to seek colors via online resources and tools.
In particular, we suggested three widely-used websites: Adobe Color~\cite{adobecolor2020}, ColorBrewer~\cite{colorbrewer2020}, and Coolors~\cite{coolors2020}. However, users could still employ any other online resources.
With these websites, users could explore numerous expert-designed or automatically generated palettes, craft palettes from scratch using harmony rules, and search palettes with words like \textit{lucky, moonlight}, covering a range of functions offered in \name{}.
We did not choose any commercial tool such as Adobe Illustrator as the baseline, because the learning curve is quite high for general users and the interfaces are dramatically different.

\textbf{Participants and Apparatus.}
We recruited 24 participants (10 females and 14 males; aged 19--26) from a local university. They are all with normal color vision and their backgrounds range from engineering, law, to business.
From a pre-study questionnaire, their average years of experience in visualization or design is 0.375 ($\sigma=0.77$), so that they are novice users for our study tasks. 
Also, their self-reported expertise of color theories (\eg, harmony rules, color semantics) was: $M=2$ and $IQR=2$, on a 7-point Likert scale (1=``do not know at all'' and 7=``very familiar'')
We deployed \name{} and its baseline version on the cloud, and participants completed the study remotely via video conferencing software on their own machines.

\textbf{Tasks.}
We created four experimental infographics with certain contextual information (\eg, talking about a kid's weekend).
Participants needed to complete two tasks during a study session. 
Task 1 aimed to assess the efficiency of the tool (S1), in which participants needed to color three infographics (out of the four) until they were satisfied with the results, one by one, without a time limitation. 
For each infographic, according to the context, three forms of color preferences were specified for three elements during the tasks, including: an exact color, a color name, and a semantic or affective word.
Task 2 aimed to assess the creativity supported by the tool (S2), in which participants colored the same infographic (the rest one of the four) within 15 minutes to obtain as many satisfying results as possible.  
In this task, users were given general contextual information instead of concrete preferences.
In each task, we explicitly explained the constraints or context and asked participants to ensure the pre-defined preferences were met and each element was distinguishable.
For the baseline, participants could import a color palette as a whole to minimize the effort of copying and pasting single colors from the websites.

\textbf{Design and Procedure.}
We employed a between-subjects design, with 12 participants finishing two tasks in each condition: \name{} or Baseline. 
We ensured that each infographic appeared in Task 2 three times in each condition across participants, and counterbalanced the order of the remaining three infographics for Task 1.
Each study session began with a tutorial about the tool (\ie, \name{} or Baseline with websites).
Then, participants completed a training task on a different infographic (than the four) with similar task requirements. They could ask any questions about the tool. 
After, participants were instructed to perform Task 1 and then Task 2 in order. They took a short break between the two tasks.
In the end, they filled in an exit-questionnaire (on a 7-point Likert scale where 1 is ``strongly disagree'' and 7 is ``strongly agree'') and the Creativity Support Index questionnaire \cite{cherry2014quantifying}, followed by a semi-structured interview.
For participants in the Baseline condition, we also briefly demonstrated \name{} and asked for their comments.
Each study session lasted around 1.5 hours and each participant received \$12 in compensation.

\begin{figure}[!tb]
    \centering
    \includegraphics[width=\linewidth]{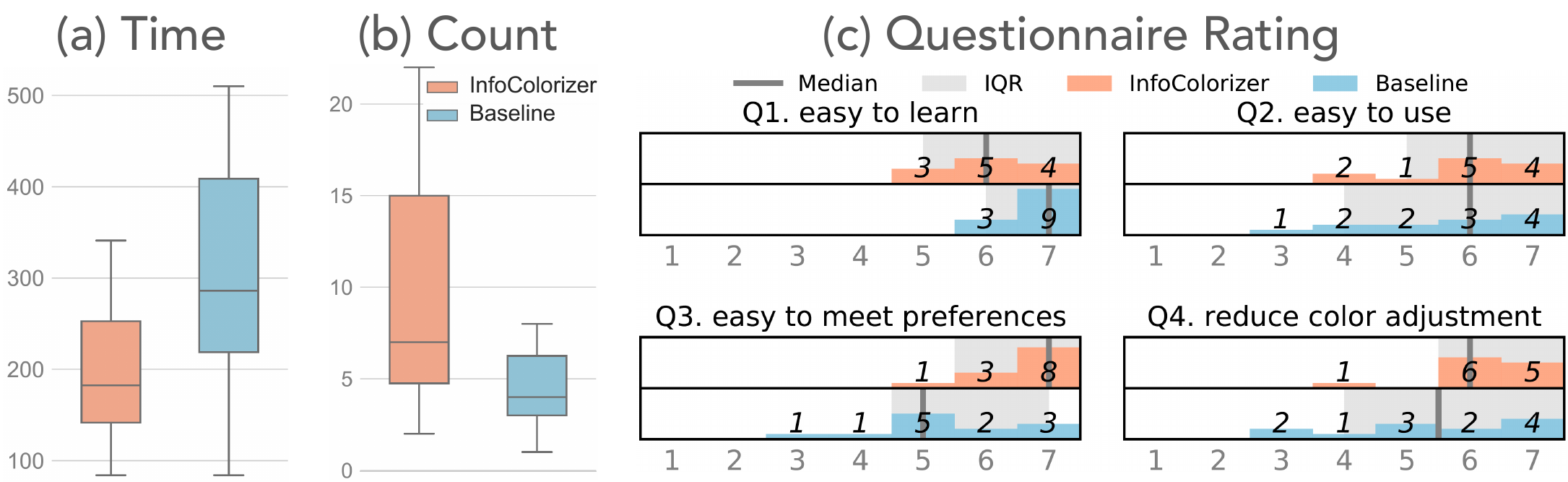}
    \vspace{-6mm}
    \caption{Results of the controlled user study: (a) Completion time for generating satisfying infographics in Task 1. (b) Counts of the resulting infographics in Task 2. (c) Participants' subjective ratings on the exit-questionnaire (the higher the better).
    }
    \label{fig:controlled_user_study}
\end{figure}

\subsubsection{Results and Analysis: Task Performance}
\autoref{fig:controlled_user_study}-a and \autoref{fig:controlled_user_study}-b show the completion time of Task 1 and the resulting infographics count in Task 2, respectively.
An independent-samples t-test showed significant differences on both the completion time ($t=-4.83, p \ll 0.001$) and infographic counts ($t=2.29, p=0.02$). 
This indicates that participants generated satisfying palettes for the infographics faster with \name{} ($\mu=197$ seconds, $95\%~CI=[173,222]$) than with Baseline ($\mu=306$ seconds, $95\%~CI=[265,346]$) for Task 1. 
Moreover, for Task 2, participants created more satisfying infographics with \name{} ($\mu=9.25$, $95\%~CI=[5.7,12.8]$) than with Baseline ($\mu=4.7$, $95\%~CI=[3.4,5.9]$).  

\subsubsection{Results and Analysis: Participants' Feedback} 
To further answer S1, besides the quantitative task performance, we analyzed participants' ratings on the exit-questionnaire ( \autoref{fig:controlled_user_study}-c).
We performed a Mann-Whitney test to compare the two conditions on each question. 

For Q1, participants generally thought Baseline was easier to learn ($U=37.5, p=0.014$), which might be because the interface of \name{} was more complicated with the recommendation function. 
Nevertheless, all participants gave a positive rating ($\geq 5$) for \name{}.

There was no significant difference on Q2 for the two conditions ($U=63.5, p=0.32$).
However, \name{} in general received more positive ratings. 
This could be because \name{} is equipped with the recommendation that benefited novices by \q{reducing the search space} (P6).
After being demonstrated \name{}, P1 from Baseline condition commented: \q{It is exactly what I want, because it can alleviate my burden to collect, assign and adjust colors.}

A significant difference was found ($U=31.5, p=0.007$) between the two conditions for Q3, indicating that \name{} allows participants to easily colorize infographics under specific color constraints.
Among the participants using the Baseline, P10 complained: \q{Sometimes I cannot get proper colors meeting preferences while looking harmonious from the three websites, so I need to determine colors in a trial-and-error process myself.} 
Additionally, P12 said that \q{I use ColorBrewer in my daily life, but the choices are limited. I also feel inconvenient when I want to change a single color in a certain palette, because the remaining colors cannot be updated.} 
Participants from the \name{} condition reported that they also encountered such difficulties; however, they felt that \name{} \q{integrates knowledge about colors, semantics and emotions,} (P7) and \q{the recommendation can solve these difficulties.} (P8).

Though no significance was found ($U=49.5, p=0.091$) on Q4, \name{} ($\mu=6.25, \sigma=0.87$) had a higher average score than Baseline ($\mu=5.42, \sigma=1.51$) and a smaller variance on Q4, indicating its stability in reducing the burden for color adjustment.
P17 using Baseline said: \q{I had no idea whether a palette would work well for an infographic after being applied, even though the palette looked fine on the websites. Thus, I needed to keep trying or finished colorizing based on my intuition.} This was also echoed by P1 and P12 (in the Baseline condition). 

\subsubsection{Results and Analysis: Creativity Support} 
To answer S2, we utilized the Creativity Support Index (CSI) \cite{cherry2014quantifying}, which quantifies how well a tool can support users' creativity based on a research-tested questionnaire. 
One usage of CSI is to compare two tools used by people from two groups for the same tasks, which is well-suited to our study.
Followed the practices in the previous work (\eg, Playful Palette~\cite{shugrina2017playful} and Color Builder~\cite{shugrina2019color}), we asked users to rate the corresponding study system on five factors: Expressiveness, Exploration, Results Worth Effort, Immersion, and Enjoyment.
\autoref{tab:CSI_scores} shows each factor's scores, based on which we calculated the final CSI scores.

\begin{table}[!tb]
    \small
    \centering
    \caption{Participants' scores on the five factors and the final CSI scores (the higher the better) of \name{} (I) and Baseline (B), with independent-samples t-test results. }
    \vspace{-4mm}
    \begin{tabular}{l|cc|l}
        \toprule
         \textbf{CSI Dimension} & \textbf{I} & \textbf{B} & \textbf{T-test} \\
        \midrule 
         Expressiveness & 33.8  & 26.7 & $t=2.84, p=0.009$  \\
        
         Exploration & 35.2  & 26.5 & $t=3.38, p=0.003$ \\
        
         Results Worth Effort & 36.3 & 25.8  & $t=9.71, p \ll 0.001$ \\
         
         Immersion & 32.5 & 27.2  & $t=2.08, p=0.049$ \\
        
         Enjoyment & 35.8 & 29 & $t=3.85, p=0.001$ \\
         \midrule 
         CSI & 57.9 & 45.1 & $t=4.75, p=0.0001$ \\
        \bottomrule
    \end{tabular}
    \label{tab:CSI_scores}
\end{table}

Overall, \name{} received a mean CSI score of 57.9 ($\sigma=6.51$), much better than that of Baseline: a mean of 45.1 ($\sigma=6.72$).  
A independent-samples t-test showed a significant difference ($t=4.75, p=0.0001$), indicating \name{} performed considerably better than Baseline.
Moreover, \name{} significantly outperformed Baseline on all the factors. The biggest difference existed in Results Worth Effort.
Though some recommended palettes were inferior to their expectation, participants still thought the overall recommendations were acceptable and the palette refinement workflow of \name{} was smoother than Baseline.
Thus, they could obtain qualified and satisfactory palettes for an infographic with less effort.
The experience in Task 2 influenced how participants felt about the Expressiveness and Exploration of \name{} and Baseline, because they needed to be creative to provide as many solutions as possible.
Several participants reported that \name{} sometimes surprised them by recommending palettes of different styles that they had never thought about, so that they could explore many different possibilities.
On the contrary, P7 using Baseline said that \q{The websites can help me search colors when I have an idea, but they cannot provide me with ideas.}
To further improve Expressiveness, P16 and P17 suggested to include more vague descriptive words into \name{}.
The better performance on Immersion and Enjoyment of \name{} may be related to the fact that it has a higher degree of integration than Baseline. This was because participants could find colors under specific preferences and assign them to elements within a single system instead of several websites.

\subsubsection{Results and Analysis: Participants' Behaviors}
We qualitatively investigated participants' behavioral patterns based on our observation, revealing that they adopted different ways to get started, obtain the first palettes, modify colors, and adjust color assignment. 

\textbf{Baseline.} 
For Task 1, it was rare for participants to use a complete palette directly from the online sources~\cite{adobecolor2020, coolors2020, colorbrewer2020}, since none could meet all the color preferences. 
Most participants started with elements having the constraints.
We noticed that Coolers~\cite{coolors2020} was more popular for them to get started.
The reasons might be: 1) compared to ColorBrewer~\cite{colorbrewer2020}, Coolers support searching palettes with keywords; and 2) Coolers has more flexible interactions to generate palettes with gradient and with over five colors than Adobe Color~\cite{adobecolor2020}. 
After coloring these elements, participants usually chose colors from the color picker and swatch embedded in the system for the remaining elements.
The swatch consisted of palettes pre-loaded from ColorBrewer and imported by them from Coolor and Adobe Color previously.
The adjustment of color palettes and color assignment happened in various ways, including: 
1) participants might adjust colors both before and after obtaining the first complete palettes; 
2) some finished quickly only via one to two adjustments, while others repeatedly assigned several colors to one element and compared the corresponding results; 
and 3) they usually either focused on elements in an arbitrary order or just from left to right, and they tended to adjust elements locally if the elements were clustered visually.

For Task 2 without requirements, they relied more on complete palettes in the provided tools by changing one to three colors or just trying different color assignments.
When searching palettes in Adobe Color and Coolors, some participants only focused on whether palettes were aesthetic while others might filter palettes using keywords fitting the infographic topic.
In both tasks, we observed that all participants did not use other online coloring tools, nor did they use the color wheel and harmony rules provided by Adobe Color.
It might be because they were novice users and unfamiliar with other tools and the rules.

\textbf{\name{}.} 
Users behaved much more consistently than Baseline on the two tasks. They started with specifying preferences by inputting words, getting recommendations, and bookmarked palettes they liked.
They might adjust one to two unsatisfactory colors by using the color picker, swatch, or recommendation functions.
More participants requested recommendations again since it allowed them to obtain many possible results.
They often finished one infographic in Task 1 within two requests. As for Task 2, they usually got the first satisfactory result within two requests and had bookmarked two to four palettes, from which they could derive more palettes in the next request. We observed that they had less hesitation and pauses during the creation than participants using Baseline. It might be because that \name{} provided them with a more consistent workflow, reduced their mental effort, and allowed them to focus on points of interest.

\subsection{Survey Study} \label{sec:survey}
Our controlled study validated the effectiveness of \name{} from an infographic creator's perspective. But how good are the generated infographics from a consumer's perspective? 
To answer this question, we conducted a survey study to evaluate the quality of \name{} recommended color palettes comparing against a set of other methods, including both human- and machine-generated palettes.
Specifically, we compare five conditions: 1) artist-designed, 2) \name{}-recommended, 3) Baseline-crafted, 4) ColorBrewer-predefined, and 5) randomly-generated color palettes. 

\subsubsection{Study Setup.}
We used the four experimental infographics mentioned in \autoref{sec:controlled_study} and crafted palettes for them under each of the five conditions.
To obtain artist-designed palettes, we asked a professional designer to create a color palette for each infographic.
For \name{} and Baseline conditions, we utilized the results generated by participants in Task 2 of the controlled user study. 
This is because no concrete preferences were set in Task 2, and thus the results were produced under the same settings among the artist-designed, \name{}, and Baseline conditions.
For the ColorBrewer condition, only categorical palettes were considered. We randomly selected a palette and assigned colors in it to the infographic elements. Lastly, we randomly generated color palettes and then randomly applied them to the infographic.
In total, for each experimental infographic, we generated one palette for artist designed condition and nine palettes for the other four conditions.
We formulated the study as online surveys.
Each survey has four problem sets, each containing five pictures derived from an identical experimental infographic but colored with five palettes, each for one of the above conditions. 
The artist designed palette was repeated across all surveys; for the other four conditions, the color palettes were randomly selected from the corresponding generated ones above.
Therefore, each survey contained $4 \times 5 = 20$ pictures in total. 
For each problem set, we asked participants to provide two 7-point scores (the higher is better), on color aesthetics and color readability. 
We randomized the order of conditions within each problem set, as well as the order of the problem sets.

\subsubsection{Participants.}
We released the survey on Amazon Mechanical Turk, and collected 102 responses in total, all valid. 
Their demographics information is as follows: 81 males and 21 females, aged 17--57 ($\mu=32.3,\sigma=8.5$), 0--16 years ($\mu=2.2,\sigma=3.5$) of experience in visualization or design, and all with normal color vision. 
Their backgrounds included science, business, finance and engineering.
Participants completed the study on their own machines.

\begin{figure}[!tb]
    \centering
    \includegraphics[width=\linewidth]{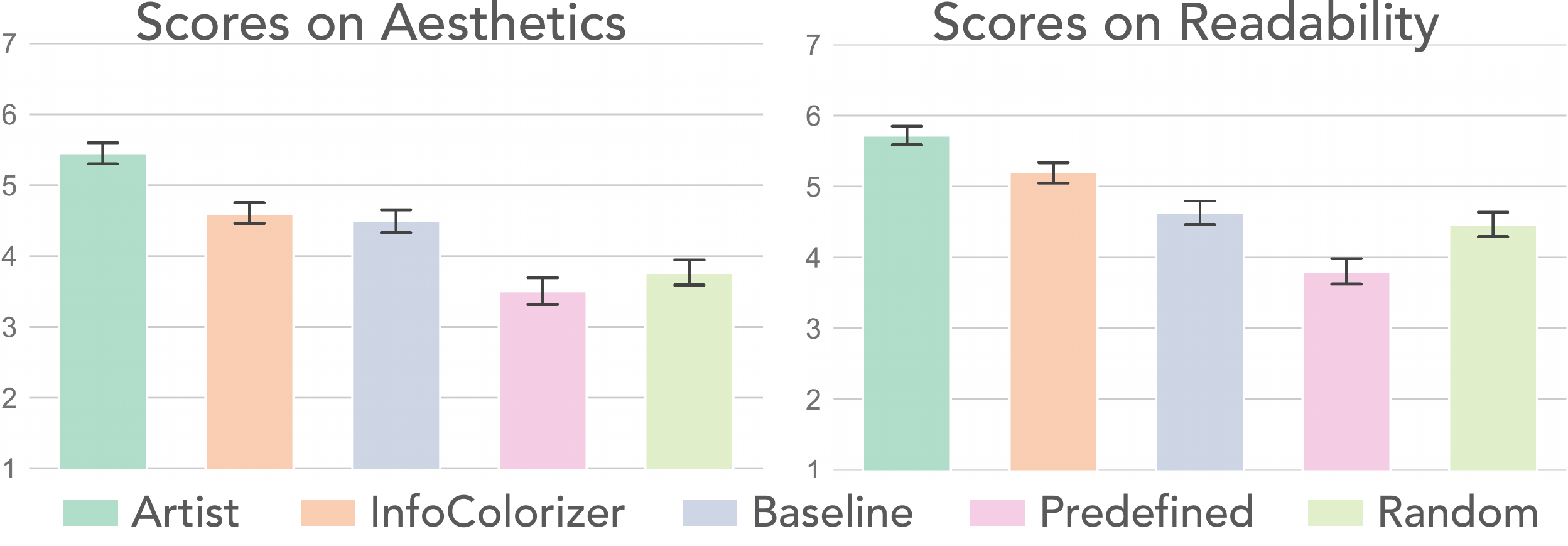}
    \vspace{-6mm}
    \caption{Survey study results: the average scores on aesthetics and readability with 95\% CIs (the higher is better; from 1 to 7).
    }
    \label{fig:survey_results}
\end{figure}

\subsubsection{Results and Analysis.} 
\autoref{fig:survey_results} shows the scores of five conditions on color aesthetics and element readability. 
We can see that artist-designed palettes received the highest average scores on both color aesthetics ($\mu=5.45, 95\%~CI=[5.31,5.59]$) and readability ($\mu=5.72, 95\%~CI=[5.59,5.85]$), respectively. 
This is plausible because these color palettes were carefully designed by the professional designer.
Among the rest, \name{}-recommended palettes obtained the highest overall scores in aesthetics ($\mu=4.60, 95\%~CI=[4.44,4.75]$) and readability ($\mu=5.20, 95\%~CI=[5.06,5.34]$).
A Friedman test indicated significant differences between different methods for aesthetics ($\chi^2= 261.27,p<0.001$) and readability ($\chi^2= 267.6,p<0.001$).
Post-hoc Dunn tests with Bonferroni adjustment showed that the score of \name{}-recommended palettes was significantly higher than that of randomly-generated color palettes both on aesthetics ($z=6.64, p < 0.001$) and readability ($z=6.11, p < 0.001$), as well as significantly higher than that of ColorBrewer-predefined palettes both on aesthetics ($z=8.53, p < 0.001$) and readability ($z=10.96, p < 0.001$).
It is not surprising that \name{} outperformed the random condition because \name{} had learned good practice embedded in expert-designed infographics while random palettes were sampled from the whole color space.
Before study, we thought that ColorBrewer might be better than \name{} on aesthetics but worse on readability.
However, the survey results show that \name{} outperformed in both aspects, indicating that even a good predefined palette can have poor aesthetics and readability because of improper color assignment.
While there was no significance between \name{}-recommended and baseline-crafted palettes on aesthetics ($z=0.69, p=0.49$), the average score of \name{} was significantly higher readability than that of Baseline ($z=4.75, p < 0.001$).
This indicates that \name{} not only improved users' creation efficiency as shown in \autoref{sec:controlled_study}, but also ensured that the recommended palettes had at least the same level of quality as user-crafted palettes with the Baseline on both aesthetics and readability.
\subsection{Interview Study}
The aforementioned studies evaluated \name{} on different aspects, including promising cases in our dataset, infographic creators' efficiency and experience compared to a baseline, and the quality of recommended color palettes based on infographic consumers' opinions. 
The users were all novices in graphics design, which belong to our target user group (\ie, general audience).
But how does the tool look like in the eyes of an expert?
Thus, we further conducted an interview study to collect in-depth qualitative feedback from graphic designers, in which they used \name{} to design color palettes for infographics of their own interests, in a more realistic setting.

\subsubsection{Study Setup}

We recruited four experts (two females and two males) from an online design community. All of them have over three years of experience in designing user interfaces, infographics, and data visualizations.

Each study session lasted about an hour and started with a 10-minute survey about their experience in design, the tools or methods they used to choose or design color palettes, and the difficulties they encountered when using those tools or methods.
Then, we provided a short tutorial of \name{} and asked the participants to freely explore the system and ask questions (20 minutes).
After, the participants completed a design task (20 minutes).
Finally, they were asked to provide ratings in a questionnaire about each system's functionality (10 minutes).
We encouraged the participants to think aloud and provide feedback and suggestions during the study. An experimenter was responsible for answering questions and taking notes.

\begin{table}[!tb]
    \small
    \centering
    \caption{Participants' ratings in the interview study.}
    \vspace{-4mm}
    \setlength{\tabcolsep}{1.5pt} 
    \renewcommand{\arraystretch}{1.2} 
    \begin{tabular}{p{7cm}cccc}
        \toprule
         & \textbf{P1} & \textbf{P2} & \textbf{P3} & \textbf{P4} \\ \midrule
         \textbf{Q1}  Tool is easy to learn & 6 & 6 & 6 & 6\\
         \textbf{Q2}  Tool is easy to use & 7 & 6 & 4 & 5\\
         \textbf{Q3}  Infographics element layers panel is intuitive & 6 & 7 & 6 & 6\\
         \textbf{Q4}  Specifying color preferences is useful & 7 & 7 & 7 & 7\\
         \textbf{Q5}  Iteratively refining the results is useful & 6 & 7 & 6 & 7\\
         \textbf{Q6}  Recommended palettes look good & 6 & 4 & 5 & 5\\
         \textbf{Q7}  Tool makes starting to design a palette easier & 6 & 7 & 6 & 7 \\
         \textbf{Q8}  Tool makes the design process more effective & 6 & 5 & 6 & 7\\
        \bottomrule
    \end{tabular}

    \label{tab:model-evaluation}
\end{table}

\subsubsection{Results and Analysis}

All participants were able to complete the design task using \name. Participants' ratings are shown in \autoref{tab:model-evaluation}. In the following, we discuss detailed results based on the questions.

\textbf{Ease of Learning and Use (Q1, Q2).}
All participants found \name{} very easy to learn, as P1 explained \q{the tool is designed similar to tools I regularly use... layers and panels are similar to document or photo editors.} Similarly, P2 added that \q{the button icons are intuitive and self-explainable.}
The responses for ease of use were mixed. While P2 gave the highest rating and thought \q{the functions are pretty intuitive.} P3 had a neutral feeling and commented that \q{some of the buttons are new to me... I do not know what they do without explanations.}

\textbf{Infographic Elements Visualization (Q3).}
All participants found the visualization of infographics elements very intuitive and useful. For example, P1 gave the highest score, commenting \q{this looks familiar to me, just like the layers panel in Photoshop} and P2 gave similar feedback that \q{If you are an Adobe user, you will understand this immediately.} 
One limitation was that the visualization is \q{missing the vertical spacial order}, as P1 explained: \q{Sometimes it is hard to locate elements if they are at the same horizontal location.} 
P1 and P2 also suggested supporting row selections, as P2 commented \q{I hope I can select layer by layer... I usually select or edit a layer a time to bind or assign them the same color.}

\textbf{Color Preference Setting (Q4).}
All participants gave the highest ratings for the Color Preference section, as P2 commented \q{It is easy to use to input my preferences} and P4 complimented that \q{Easy to rapidly adjust color... sufficient for most of my needs.} 
Specifically, P1 liked the combination of manual (i.e., color assignment) and automatic (i.e., vague preference) methods and said \q{the functionalities for specifying colors are very complementary to each other... some are manual but can see effects immediately... some are automatic but need to re-run the model.} On the other hand, P4 particularly liked the color binding feature and explained \q{It allows users to input the relationships among the elements to the model... the model won't be able to know this information without human input.}
Beyond the controls, P2 suggested that \q{it will be amazing if it can learn and memorize users' color preferences from their design history}
P3 and P4 suggested adding some explanations to the vague color specification.

\textbf{Recommendation Quality (Q6).}
Most participants thought the quality of the recommended color palettes was good, as P1 applauded that \q{the color contrast between the background and foreground is good.} 
P2, who gave a neutral rating explained that \q{I cannot see clear color contrast between layers.} She suggested a solution that \q{Maybe we should have some rules to force it... it would be great if I can specify my vague preference for each layer.} 
Two participants also suggested grouping the recommendations by similarity, as P4 explained that \q{I hope the model can group similar palettes together, so I only need to try one from each group to quickly find out which style is the best.}
In addition, P3 asked for more explanations behind each recommendation: \q{Suppose the system is already learning from professional designers, maybe it can explain the recommendation by telling me the styles or names of the designers, so I can learn their styles. Great for non-expert designers to improve their skills by learning.}

\textbf{Iterative Workflow (Q5, Q7).}
Overall, all participants thought \name{} made it easy to get started with designing a color palette and that the iterative refinement process was effective for producing high-quality designs. 
For example, P2 commented that \q{Usually I don't know what color I want at the beginning... so having some recommendation is helpful to get started} and P3 added \q{I like having a diverse set of options to explore at the beginning.}
P1 found the iterative refinement process very effective and explained that \q{compared to the original designs, the initial recommendation already looks good but a little bit random... after I specific rules, it narrows down the design spaces and starts to give more personalized recommendations.} 
Similarly, P4 also reflected on her design iterations and commented that \q{at beginning, the search space is usually large... with the recommendations, I only need to review a few to identify what styles fit the best, so I can quickly reduce the search space.}
To improve the design process, P1 hoped \name{} \q{can memorize my history, so the system will become smarter even at the first recommendation.} 
P3 pointed out a limitation that \q{the recommendations tend to be similar to what I selected in the last round, even though I did not want to}. 

\textbf{Comparison to Existing Tools (Q8-11).}
During the survey, the participants introduced the existing tools they used for choosing or designing color palettes. These tools can be grouped into three categories: 1) manual, such as assigning colors in Photoshop or Illustrator, and 2) half-manual, such as tools recommending high contrast colors to a specified color, and 3) templates, such as themes in PowerPoint.
Compared to existing tools, most participants thought \name{} makes the palette design process more effective since it is more automatic and personalized. 
Compared to the template tools, P2 commented that \q{it (\name{}) is more flexible than pre-defined themes... I can specify color preferences and choose from a large set of options} and P3 added that \q{pre-defined themes are not considering the specific design I want to make... it always give you the same set of templates and makes your design look similar to others.} 
Compared to the manual or half-manual tools, P4 thought \name{} is more efficient since \q{Manual is not scalable... I can process more designs with the recommendation workflow.}
All participants mentioned that \name{} can be used along with their existing tools. P1 explained in details that \q{Here (\name{}) we are starting from scratch. I hope to start from some pre-defined themes... It is more a trade-off between quality and efficiency for different use cases and scenarios.}

\section{Discussion}
Here we discuss several aspects about \name{} and our studies, including limitations and potential solutions.

\textbf{Generalization for Different Infographics.}
Our current recommendation engine is designed for infographics that can be described by the conceptual model in \autoref{sec:conceptual-model}.
However, there are some percentage of infographics containing data charts (\eg, line charts, scatterplots).
We observed about 1\% of such infographics in InfoVIF.
While not a big percentage, \name{} may fail to recommend proper palettes for these infographics, because embedded data is not characterized in input features.
This can be addressed by collaborating with tools for data charts~\cite{harrower2003colorbrewer, gramazio2016colorgorical}. For example, to colorize an infographic with a bar chart, a user can first obtain colors from Palettailor~\cite{lu2020palettailor}, assign them to each bar as color preferences, and get recommended colors for the rest elements with \name{}. 
This process may be further automated by integrating prior work on colorizing data charts \cite{wang2018optimizing, lee2012perceptually} as the conditions of VAEAC.

Our method may also not work well on infographics with complicated clipart images, which require advanced computer vision techniques to recognize and segment objects. 
Many of the images serve as a semantic background that may influence the color appearance of the whole infographic.
However, when suitable techniques are available, the information of objects in images can be integrated into our tree model, and thus we can still apply our data-driven method to learn design practices and recommend palettes.  

\textbf{Explicit and Implicit Color Constraints.}
When recommending palettes, we only consider color preferences explicitly assigned by users, and leave aside possible implicit constraints exhibited in infographics.
For example, if a sequential palette is used for encoding data in an infographic, our generated palettes may fail to remain the relationship.
Currently, a user has to specify the sequential relationships using exact colors or vague words (\eg, in \autoref{fig:case_study}-B3).
A solution can be embedding such relative color relations (\eg, sequential) into the feature vector, and train VAEAC to learn these patterns.
Similarly, as infographics can benefit from using gradient colors, integrating gradient colors into the features would be interesting to explore.

\textbf{Generalization and Penalization Trade-off.}
Our method is data-driven, meaning that the style and quality of recommended palettes depend on the training data. 
More training data will likely enhance the model with generality and accuracy. 
However, whether a palette is aesthetic or not is still subjective.
Currently, we choose VAEAC which can generate diverse, as well as relevant, palettes (Appendix Sec. B) to accommodate users with different aesthetic tastes.
One solution can be training a more personalized model gradually based on the resulting palettes chosen by a user.

\textbf{Limitations in Study Design.}
For the survey study, we randomly assigned colors to infographic elements for the ColorBrewer condition. 
Without manual adjustment, adjacent elements may be assigned with the same color, reducing the readability. 
However, this is a common situation in real world; and one of the challenges that we address here is the color assignment problem. 
Further, we note that the sample size of our interview study might be small. 
However, we obtained deeper insights regarding \name{} and their infographic creation workflow, and our controlled study with more users complements this effect to some extent.
But a future deployment study may be needed to evaluate the usefulness of \name{} with more realistic settings for a long term.

\section{Conclusion and Future Work}
We have introduced \name{}, an interactive system that supports effective infographic color palette design via cogent recommendations.
The system leverages a conceptual infographic model and deep learning techniques to lower design barriers, support flexible color preference specification, and adapt palettes generation to spatial relationships of infographic elements. 
We have demonstrated the effectiveness and usefulness of \name{} through case studies, a controlled user study, a survey study, and an interview study.
Our work opens several avenues for future work. 
We plan to explore metrics to rank returned palette recommendations, which can further reduce users' effort to examine and choose palettes.
We also would like to support more advanced color preferences such as relative lightness and perceptual differences between two elements.



\bibliographystyle{abbrv}
\bibliography{recolor}

\clearpage
\appendices
\section{Feature Explanation and Extraction}
\label{apx:features}
\subsection{Feature Explanation}

As mentioned in \autoref{sec:features}, we distill a list of features to characterize an infographic at multiple levels. Below we give a detailed explanation of each non-color feature and illustrate them with the infographic shown in \autoref{fig:art-element}.

\textbf{Infographic Level.} We use the following features:
\begin{itemize}[leftmargin=1em, itemindent=0.0em, nosep]
    \item \textit{VIF Type} is the underlying narrative structure (visual information flow) of an infographic \cite{lu2020exploring}, where there are 12 types of VIF, such as \textit{Landscape}, \textit{Portrait}, \textit{Clock}, \textit{Up-ladder}. The VIF type of \autoref{fig:art-element} is \textit{portrait}.
    
    \item \textit{Visual Group Number} is the number of visual groups on the VIF backbone. There are two visual groups (the first A1 and the second B2 row) in \autoref{fig:art-element}.
    
    \item \textit{Visual Group Distance} is the average distance between the centroids of two adjacent visual groups on the VIF backbone. The distance between the two groups in \autoref{fig:art-element} can be calculated as the distance between centers of two circles (Element 3 and 8).
\end{itemize}

\textbf{Visual Group Level.} We consider the following features:
\begin{itemize}[leftmargin=1em, itemindent=0.0em, nosep]
    \item \textit{Visual Group Element Number} is the number of (artistic and graphical data) elements within a visual group. In \autoref{fig:art-element}, each group has eight elements.
    \item \textit{Relative Visual Group Size} is the width and height of the bounding box of a visual group divided by the width and height of the infographic image, respectively.
\end{itemize}

\textbf{Element Level.} We extract the following features for each artistic and graphical data element:
\begin{itemize}[leftmargin=1em, itemindent=0.0em, nosep]
    \item \textit{Element Type} classifies the appearance of an element, where for an artistic element, it can be \textit{triangle}, \textit{square}, \textit{rectangle}, \textit{pentagon}, \textit{circle} or \textit{others}, and for a graphical data element, it can be \textit{index}, \textit{text}, \textit{icons} or \textit{arrows} \cite{lu2020exploring}. In \autoref{fig:art-element}, A1 and B2 are text, and their background shapes are pentagons.
    
    \item \textit{Relative Element Size} is the width and height of its bounding box divided by the width and height of the infographic, respectively.
    
    \item \textit{Relative Element Pixel Area} is the pixel area of an element divided by the total pixel area of the infographic. Note that the pixel area of an element is not necessary the same as its bounding box (\eg, text, icons, index, and non-convex shapes).
\end{itemize}

To represent spatial arrangement within an infographic,  we adopt the nested set model \cite{feichtner2005topology} to traverse its corresponding tree structure described in \autoref{sec:conceptual-model}. 
In particular, we store the following information of each node:
\begin{itemize}[leftmargin=1em, itemindent=0.0em, nosep]
    \item \textit{Left Index Number and Right Index Number} of a node are the visiting sequence numbers generated in a pre-order traversal where each node is visited twice and thus two indices are assigned.
    Every tree structure is then uniquely associated with these left and right node index numbers.
\end{itemize}

\subsection{Technical Details for Features Extraction}
While Lu \etal \cite{lu2020exploring} provided methods for extracting VIF and graphical data elements, our key technical challenges include identifying artistic elements and constructing the tree structure as described in \autoref{sec:conceptual-model}.

\textbf{Infographic Level Features Extraction.} 
To get these features, we employ the data element extraction and VIF construction algorithms in \cite{lu2020exploring}. 
Their data element extraction utilizes the state-of-art object detection model, YOLO \cite{redmon2016yolo}, to identify the bounding boxes of graphical data elements (\eg, icons, text, indices) in an infographic. 
Based on the detected elements, the VIF construction algorithm leverages Gestalt principles (\eg, proximity, similarity, and regularity) to identify the visual groups and VIF backbone.
Therefore, we can easily compute the \f{VIF Type}, \f{Visual Group Number}, and \f{Visual Group Distance}.

\textbf{Artistic Elements Identification.}
The algorithms in \cite{lu2020exploring} can only detect graphical data elements, whereas identifying artistic elements is essential for us to compute the features at the visual group and element levels and to construct a precise tree model of an infographic.
An intuitive idea is to find areas with the same (or similar) colors using color segmentation \cite{tremeau1997region}, because an artistic element is usually exhibited as a shape with a consistent color or smooth color gradient.
We achieve this via three main steps (\autoref{fig:art-element}):

\begin{figure*}[!tb]
    \centering
    \includegraphics[width=\linewidth]{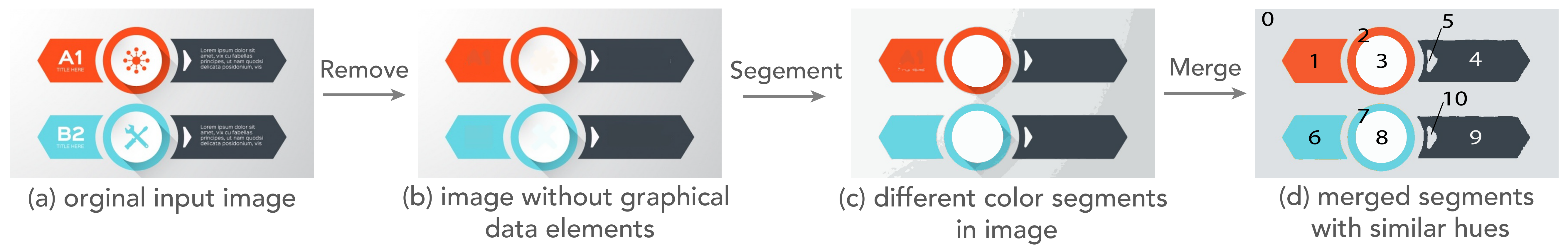}
    \vspace{-8mm}
    \caption{Identifying artistic elements in an infographic.}
    \label{fig:art-element}
\end{figure*}

\begin{enumerate}[leftmargin=1em, itemindent=0.0em, nosep]
    \item \textit{Remove graphical data elements.} 
    The graphical data elements of an infographic can interfere with our color segmentation, because they are also colored and often overlay on top of artistic elements.
    As shown in \autoref{fig:art-element}-b, to remove a data element, we simply set all the pixels within the bounding box with its background color (usually the color of the artistic element below or the infographic background) based on the results of YOLO \cite{redmon2016yolo}. 
    
    \item \textit{Perform color segmentation.}
    With the ``clean'' infographic from last step, we group continuous pixels with similar colors (in CIELab color space) using a region growing algorithm \cite{tremeau1997region} (\autoref{fig:art-element}-c).
    A threshold in CIEDE2000 color difference \cite{sharma2005ciede2000} needs to be set, and we experimentally found that setting the threshold to 4 yields good segmentation results.

    \item \textit{Merge segments with similar color hues.}
    An artistic element may contain a color gradient, which results in multiple segments from the previous step. 
    As these segments usually have similar hue, we apply kernel density estimation (KDE) clustering \cite{kim2012robust} on hue and group segments in the same cluster into one (\autoref{fig:art-element}-d).
    In particular, we used Gaussian kernel and set the bandwidth to 3.
\end{enumerate}

We therefore obtain a continuous region of pixels that represents an artistic element, and thus can easily compute its bounding box.

\begin{figure}[!tb]
    \centering
    \includegraphics[width=\linewidth]{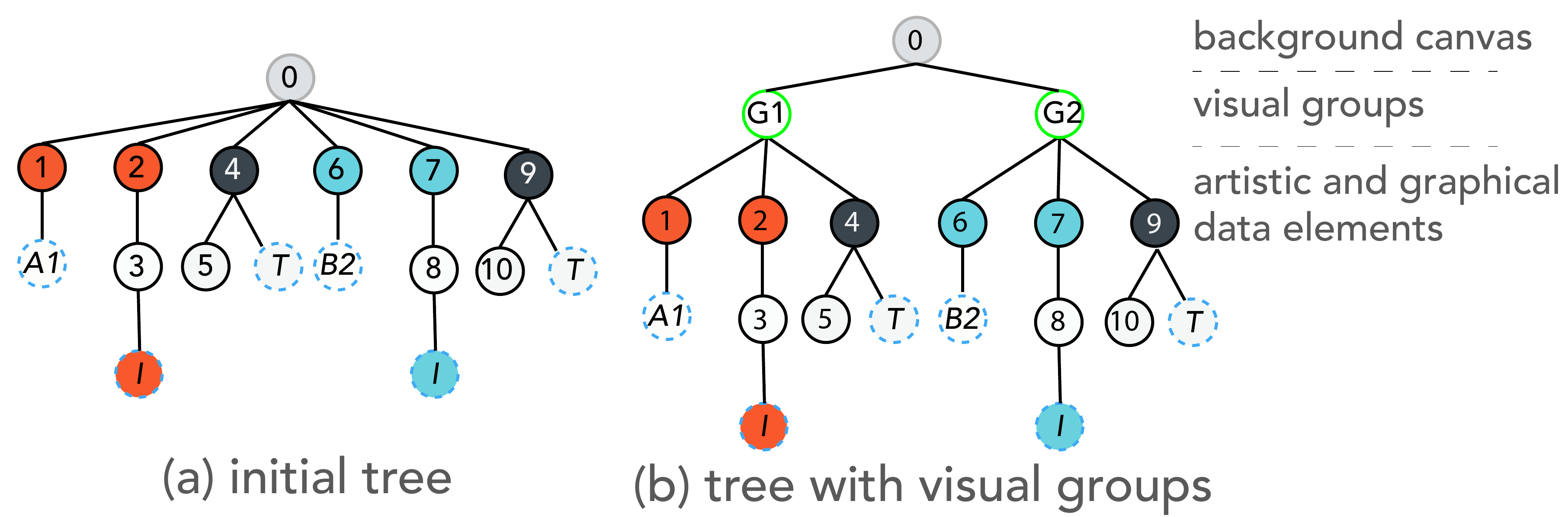}
    \vspace{-8mm}
    \caption{Constructing the conceptual tree model (see \autoref{sec:infographic-model}) with the infographic in \autoref{fig:art-element}. Black, blue, and green stroked circles indicate graphical data elements, artistic elements, and visual groups, respectively.}
    \label{fig:tree-construction}
\end{figure}

\textbf{Tree Construction.} 
With all the graphical data and artistic elements identified, we now can construct a tree structure described in \autoref{sec:conceptual-model} based on their bounding boxes. 
We start by considering each element as a node, and construct the tree from top to bottom. 
An edge is added between two elements if one's bounding box directly contains the other's without others spatially in-between. 
As shown in \autoref{fig:tree-construction}-a, we then obtain a tree whose root node is the background canvas of an infographic and other nodes are either graphical data elements (in blue dashed strokes) or artistic elements (in black strokes). 
Next, we group branches containing graphical data elements within a visual group (based on the VIF construction algorithm), and insert visual group nodes (in green strokes) below the root (\autoref{fig:tree-construction}-b).
With this tree representing the logical structure as well as the spatial arrangement of elements in an infographic, we conduct a pre-order traversal on the tree and compute the \f{Left Index Number} and the \f{Right Index Number}.

\textbf{Visual Group Level and Element Level Features Extraction.} 
After identifying all the elements and visual groups with the above procedures, most of the visual group and element level features (such as \f{Visual Group Element Number}, \f{Relative Visual Group Size}, \f{Relative Element Size}, and \f{Relative Element Pixel Area}) are easy to compute. 
The \f{Element Type} of graphical data elements can be obtained by the data element extraction in \cite{lu2020exploring}. 
To recognize the \f{Element Type} of an artistic element, we first find its contour (consisting of pixel locations) using Suzukil \etal's algorithm \cite{suzuki1985topological}; and then we compute an approximated contour (consisting of a few vertices) using the Ramer-Douglas-Peucker algorithm \cite{ramer1972}. 
We classify the artistic elements based on their shape using the vertices in the approximated contour. 
For example, if there are three vertices, we recognize it as a \textit{triangle}. 

\section{Model Training and Evaluation}
\label{apx:model_training}
\subsection{VAEAC Training}
We trained a VAEAC (Variational AutoEncoder with Arbitrary Conditioning) \cite{ivanov2018variational} model based on feature vectors $[\mathbf{F},\mathbf{C}]$ extracted from a large expert-designed infographic collection (\autoref{sec:recommendation-model}).
To obtain a fixed-length vector based on the flattened tree, we limited the maximum number of nodes as 19 based on our observation of the infographics in InfoVIF \cite{infovif}. 
Zeros were filled in the feature vector if there were not enough nodes.
The final dataset contained 8,278 infographics after removing those with more than 19 nodes. 
We split the data into 80\% for training and 20\% for testing. 
We further used 10\% of the training data as the validation set to select the best model during training.

\subsection{Model Alternatives and Evaluation}
In developing \name{}, we considered two alternative models solving the same problem as VAEAC including GAIN (Generative Adversarial Imputation Nets) \cite{pmlrv80yoon18a} and MICE (Multivariate Imputation by Chained Equations) \cite{buuren2011mice}. 
We trained a GAIN and MICE model on the same set of feature vectors $[\mathbf{F},\mathbf{C}]$ as VAEAC model. We also investigated whether the spatial features would influence the effectiveness of the VAEAC. 
To do so, we obtained new feature vectors $[\mathbf{F'},\mathbf{C}]$ from $[\mathbf{F},\mathbf{C}]$ by removing spatial features, encoded by \textit{Left Index Number} and \textit{Right Index Number}. 
We then trained a non-spatial VAEAC model based on $[\mathbf{F'},\mathbf{C}]$ with the same network architecture and hyperparameters as the VAEAC model.

To evaluate the models, we adopted a similar approach as in \cite{ivanov2018variational}. 
For each infographic in the test set, we randomly dropped 50\% of the color features $\mathbf{C}$ as the ``missing'' features; therefore, we had the ground truth information that is the original $\mathbf{C}$. 
We replaced each infographic by five different ones with random unobserved color features; thus, the test data size increased by five times.
In the experiments, for each model, we generated five full color features $\mathbf{C}$ for each test infographic.

\begin{table}[!tb]
    \small
    \centering
    \vspace{-6mm}
    \caption{Comparison of model performances with NRMSE (lower is better), Color Relevance Score (CRS, lower is better), and Color Variance Score (CVS, higher is better). 
    }
    \setlength{\tabcolsep}{6pt} 
    \renewcommand{\arraystretch}{1} 
    \vspace{-4mm}
    \begin{tabular}{cccc}
        \toprule
         & \textbf{NRMSE} & \textbf{CRS} & \textbf{CVS} \\
        \midrule 
         VAEAC & \textbf{0.6543} & \textbf{2.4826} & 5.6748 \\
         GAIN & 2.4574 & 4.1742 & 4.1075 \\
         MICE & 15.6098 & 16.5096 & \textbf{27.6199} \\
         VAEAC (non-spatial) & 1.1536 & 3.6874 & 6.429 \\
        \bottomrule
    \end{tabular}
    \label{tab:model-evaluation}
\end{table}

We considered three metrics for assessing the model performance: \textit{NRMSE}, \textit{Color Relevance Score} (CRS), and \textit{Color Variance Score} (CVS).
NRMSE is Root Mean Square Error (RMSE) normalized by the standard deviation of each feature. 
For each test case, we computed this measure via $\frac{1}{n}\sum_{i=1}^{n} NRMSE(\mathbf{C}_o,\mathbf{C}_i)$, where $n=5$, $\mathbf{C}_o$ is the original feature, and $\mathbf{C}_i$ is the imputed one.
CRS measures the degree of relevance between the ground truth and the generated color features: $\sum_{i=1}^{n} d(\mathbf{C}_o,\mathbf{C}_i)$, where $d=\frac{1}{m}\sum_{k=1}^m CIEDE(\mathbf{C}_o^k,\mathbf{C}_i^k)$. 
$CIEDE(\cdot)$ is the CIEDE2000 difference \cite{sharma2005ciede2000} between the corresponding $m$ pairs of colors, $\mathbf{C}_o^k$ and $\mathbf{C}_i^k$, in the feature vectors. 
CVS measures the degree of variance among the generated color features, which is computed by  the pairwise color differences: $\sum_{i=1}^{n}\sum_{j=i+1}^n d(\mathbf{C}_i,\mathbf{C}_j)$.
The above measures were computed for each test case, and we report the averages across the test set in \autoref{tab:model-evaluation}.
We can see that VAEAC had the lowest NRMSE and CRS while having higher CVS than GAIN.
While MICE had the highest CVS, its other two metrics were the lowest. 
We also note that the spatial features had a positive influence. Compared to non-spatial VAEAC, VAEAC had lower NRMSE and CRS. This indicates that VAEAC successfully captured the relationships between the colors and the spatial features. 

Therefore, we chose VAEAC trained with spatial features as the basis of \name{}'s recommendation engine.

\end{document}